\documentclass[reprint,superscriptaddress,aps]{revtex4-1}

\usepackage{xcolor}

\usepackage{graphicx}
\usepackage{bm}
\usepackage{hyperref}
\usepackage{color}
\usepackage{amsmath}
\usepackage{pifont}
\usepackage{epstopdf}
\usepackage{orcidlink}
\usepackage{tikz}

\hypersetup{
  hidelinks,
  colorlinks=true,
  linkcolor=black,
  citecolor=black
}

\begin{document}

\title{Electronic Phase Propagation Speed in BaFe$_2$As$_2$ Revealed by Dilatometry}
\date{today}  

\author{Xin Qin}
\affiliation{International Center for Quantum Materials,
  Peking University, Haidian, Beijing 100871, China}
\affiliation{Hefei National Laboratory, Hefei 230088, China}
\author{Xingyu Wang}
\affiliation{Beijing National Laboratory for Condensed Matter Physics,
  Institute of Physics, Chinese Academy of Sciences, Beijing 100190, China}
\author{Wenshan Hong}
\affiliation{International Center for Quantum Materials,
  Peking University, Haidian, Beijing 100871, China}
\affiliation{Hefei National Laboratory, Hefei 230088, China}
\author{Mengqiao Geng}
\affiliation{International Center for Quantum Materials,
  Peking University, Haidian, Beijing 100871, China}
\affiliation{Hefei National Laboratory, Hefei 230088, China}
\author{Yuan Li}
\affiliation{International Center for Quantum Materials,
  Peking University, Haidian, Beijing 100871, China}
\affiliation{Hefei National Laboratory, Hefei 230088, China}
\author{Huiqian Luo}
\affiliation{Beijing National Laboratory for Condensed Matter Physics,
  Institute of Physics, Chinese Academy of Sciences, Beijing 100190, China}
\author{Shiliang Li}
\affiliation{Beijing National Laboratory for Condensed Matter Physics,
  Institute of Physics, Chinese Academy of Sciences, Beijing 100190, China}
\author{Yang Liu}
\email{liuyang02@pku.edu.cn}
\affiliation{International Center for Quantum Materials,
  Peking University, Haidian, Beijing 100871, China}
\affiliation{Hefei National Laboratory, Hefei 230088, China}

\date{\today}

\begin{abstract}

  Thermal expansion offers deep insights
  into phase transitions in condensed matter physics. Utilizing an
  advanced AC-temperature dilatometer with picometer resolution, this
  study clearly resolves the antiferromagnetic and structural transition in
  BaFe$_2$As$_2$. The implementation of temperature oscillation
  reveals a hysteresis near the transition temperature $T_\mathrm{N}$ with
  unprecedented resolution. Unexpectedly, we find that the hysteretic
  width exhibits a universal dependence on the parameters of
  temperature oscillation and the sample's longidutinal dimension,
  which in turn reveals a finite transition speed. Our quantitative
  analysis shows that this phase boundary propagates at a mere 188
  $\mu$m/s – a speed seven orders of magnitude slower than acoustic
  waves. It suggests a hidden thermodynamic constraint
  imposed by the electronic degrees of freedom. Our research not only sheds light on the dynamics of phase transitions between different correlated phases, but also establishes high precision dilatometry as a powerful tool for material studies. This measurement technique, when properly modified, can be extended to studies of other material properties such as piezoelectric, magneto-restriction, elastic modulus, etc.

\end{abstract}

\maketitle

\section{\label{introduction}INTRODUCTION}

Phase transitions are of great interest in study of materials, where a variety of degrees of freedom are often coupled
together. It is important to experimentally differentiate closely
related transitions, as well as to identify the primary driving force
behind transition. Because density is a true scalar that
remains invariant under all symmetry operations relevant to solids, it
is expected to have symmetry-allowed coupling to all phase
transitions. As a result, accurately measured density can be used for
detection and classification of phase transitions. In practice, this
is often done with length measurements such as the linear thermal
expansion coefficient $\alpha$\cite{RN199, RN225, RN217, RN224, RN218,
  RN216}. Along with atomic microscope piezocantilever, strain gauge,
piezobender and X-ray diffraction\cite{RN231, RN232, RN238, RN241},
the mostly used and accurate dilatometer is based on plate
capacitors. The length change of the sample $\Delta L$ is captured
through monitoring the capacitance between the sample's upper surface
and a metal reference surface, and $\alpha$ is deduced by numerical
differentiation $\alpha=L^{-1}{dL}/{dT}$ \cite{RN226, RN227, RN228,
  RN229, RN215, RN240}. Unfortunately, the sample is always at
equilibrium, because this method needs an ultra-high temperature
uniformity of the whole mechanichal structure to achieving its high
accuracy, and its capacitance measurement limits temperature varying
speed to $\sim 10^{-3}$ K/s.

In condensed matter physics, the dynamic behavior of transitions
is always attractive and many techniques have been implemented \cite{RN242,
  RN243, RN244, RN245, RN246}. Limited either by resolution or slow
response, conventional dilatometers are not capable in studying the
dynamic procedure of phase transitions. In this work, we present a new
approach of measuring thermal expansion with pm-resolution using
optical interferometers. Using oscillating temperature, we can analyze
the dynamic response of the thermal expansion in the frequency domain.
As a demonstration, we examine the anti-ferromagnetism phase
transitions in BaFe$_2$As$_2$ \cite{RN214, RN212, RN211, RN213}. Our
systematic investigation reveals a hysteresis that originates from the
nonequilibrium state of sample near the transition temperature
$T_\mathrm{N}$. The width of the hysteresis in temperature is
proportional to the sample thickness $L$, the frequency and amplitude
of temperature oscillation $\delta T$. Thanks to the continuous
perturbation by $\delta T$, this hysteresis is likely a manifestation
of the fact that the phase transitions is not infinitely sharp in time
and the phase boundary propagates with a finite speed. The measured
speed $\simeq 188$ $\mu$m/s is significantly smaller than the acoustic
velocity, shining light on the complex nature of domain boundary
dynamics.

\begin{figure}[htbp]
\includegraphics[width=0.48\textwidth]{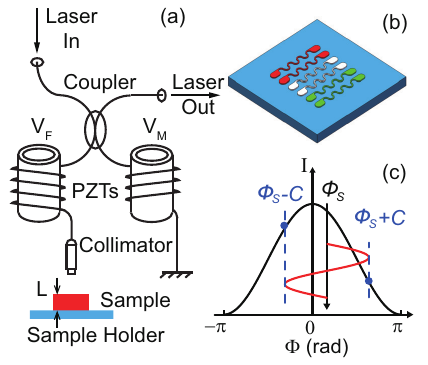}%
\caption{(a) Our dilatometer implements a fiber Michelson
  interferometer. The probing (left) and reference (right) arms are winded around
  cylindric PZT rings by which we can tune their length using
  the modulation and feedback voltages $V_M$ and $V_F$. $V_M$ generate a phase
  modulation $\phi_M=C\cos(\omega_M t)$, and $V_F$ eliminates the thermal drift $\Phi_0\approx 0$. (b) A detailed
  cartoon of the sapphire sample holder. The white, red and green
  meadow Pt wires are used as thermometer, AC and DC heaters. (c) The interfered light
  intensity $I=\cos(\Phi)$ is an odd function of the two beams'
  phase difference $\Phi=\Phi_M+\phi_S$, where $\phi_S$ origins from the thermal expansion induced by the AC temperature oscillation. $\Phi$ oscillates between $\phi_S\pm C$ and the optical power difference between the two $\Phi$ extremes has a positive dependence on $\phi_S$; see text for detailed description of our measurement principle. 
}
\end{figure}

\section{\label{EXPERIMENTAL SETUP}EXPERIMENTAL SETUP}

We use BaFe$_2$As$_2$, the parent compound of the “122”
Fe-based superconductors, an ideal prototypical example of phase transition(s) with coupled degrees of freedom, for demonstration. It 
is generally believed to have a two-step phase transition when temperature reduces
\cite{RN214, RN212, RN211, RN213}, i.e. a second-order structural
transition at $T_\mathrm{S}=134.5$ K followed by a first-order magnetic
transition at $T_\mathrm{N}=133.75$ K \cite{RN221, RN230}. An electronic nematic
phase is reported between $T_\mathrm{S}$ and $T_\mathrm{N}$
\cite{RN200, RN198, RN220, RN219, RN223}. 
We study four single-crystal as grown samples (S1 to S4) with typically in-plane size as (2$\sim$4)$\times$(2$\sim$4) mm$^2$ and different
thickness $L=$430, 210, 400 and 75 $\mu$m, respectively. Sample S1
and S2 are cleaved from the same piece, while S3 and S4 come
from another one. 

The samples are placed on upper surface of sapphire sample holder with extremely thin ($\sim$ 5$\mu$m) N-Grase for better thermal conductivity so they are considered to be free-standing and the thermal expansion is measured along their $c$-axis direction.
There are three pairs of evaporated Pt wires on the lower surface of sample holder which are used as thermometer, AC
and DC heaters, respectively; see Fig. 1(b). We measure the thermometer
resistance and calibrate it by the cryostat temperature
when the heaters are switched off. The real-time sample temperature can be separated into a
slowly sweeping DC part $T_\mathrm{0}$ by DC heater and an AC oscillation $\delta T$ by AC heater. $\delta T$ of sample induces an oscillation $\delta L$ in sample thickness and then the thermal expansion coefficient $\alpha$ can be deduced by $\alpha=L^{-1}\delta L/\delta T$ with the measurements of $\delta L$ and $\delta T$ by lock-in technique. We are able to achieve peformance comparable to the best reported capacitive dilatometry\cite{RN240}, i.e. pm-resolution in $\delta L$ amplitude $\langle \delta L \rangle $ and mK-resolution in $\delta T$ amplitude $\langle \delta T \rangle $, with dynamics measurement nature.

The $\delta L$ measurement of our dilatometer is based on a high resolution optical fiber Michelson interferometer with  and the output interference light intensity $I$ depends
on the phase difference $\Phi$ between probing and
reference light beam as $I=I_0[1+\cos(\Phi)]$. We can tune the
optical length of the two beams by applying the modulation and
feedback voltages, $V_M$ and $V_F$, to the corresponding PZT rings;
see Fig. 1(a). $\Phi$ consists three different components: the
modulation phase $\phi_M=C\cos(2\pi f_Mt)\propto V_M$, the AC phase
signal $\phi_S= 4\pi /\lambda\cdot \delta L$ where $\delta L = \alpha
L \delta T$ is the sample's thermal expansion induced by AC temperature oscillation
$\delta T$, and the slowly varying $\phi_0$ caused by the thermal drift. $C$ and
$f_M$ are the amplitude and frequency of the modulation, $\alpha$
is the thermal expansion coefficient and $\lambda=1550$ nm is the
optical wavelength. We compensate the thermal drift using the feedback
voltage $V_F$ so that $\phi_0$ can be neglected. According to the
Jacobi-Anger expansion, the amplitudes of the output optical power's $1^{\text{st}}$ and $2^{\text{nd}}$
harmonic component at $f_M$ and $2f_M$ are $I_1= I_0J_1(C)\sin(
\phi_S(t))$ and $I_2= I_0J_2(C)\cos(\phi_S(t))$, respectively; $J_N$ is
Bessel function of the $N$-th order. We measure $I_1\&I_2$ using lock-in technique
to deduce the sample's thermal expansion through
\[ \delta L=\frac{\lambda}{4\pi}
  \tan^{-1}(\frac{J_2(C)I_1}{J_1(C)I_2})\approx
  \frac{J_2(C)\lambda}{4\pi J_1(C)I_2} \cdot I_1\]
The phase modulation frequency $f_M$ is typically a few
kHz, the AC temperature oscillation frequency $f_S$ is 100 and 250 mHz,
the feedback eliminates drifts at $\lesssim$ 0.1 Hz, and the $T_\mathrm{0}$
sweeping rate is $10^{-4}$ K/s. Our resolution of $\langle\delta L\rangle$ is as small as about
5 pm when using 50 mHz resolution bandwidth, so that we can use $\langle\delta T\rangle$ as small as 5 mK, see Fig. 3. You can find more detials about the design, principle and peformance of our dilatometer in Ref. \cite{RN239} if you're interested.

\begin{figure}[htbp]
  \includegraphics[width=0.48\textwidth]{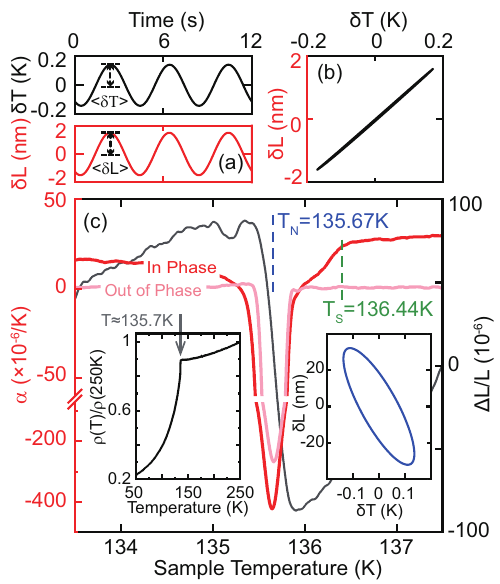}
  \caption{(a) The typical $\delta L$ and $\delta T$ oscillations measured at $T_\mathrm{0}\simeq 130$ K. (b) The $\delta L/L$ vs. $\delta T$ plot of panel (a) data. (c) The in-phase (red) and out-of-phase (pink) thermal expansion coefficient $\alpha$ by comparing $\delta L$ and $\delta T$ oscillations. The DC thickness variation $\Delta L$ (black) is deduced by $G\cdot V_F$. Data measured from sample S1 using $\delta T$ frequency $f_S=250$ mHz and amplitude $\langle \delta T\rangle =0.14$ K. The magnetic transition temperature $T_\mathrm{N}$ and structural transition temperature $T_\mathrm{S}$ are marked by blue and green dash lines, respectively. The left inset is the zero-field normalized resistivity in (ab) plane and the right inset illustrates the $\delta L$ vs. $\delta T$ hysteresis loop at $T=T_\mathrm{N}$.}
\end{figure}

\section{\label{EXPERIMENTAL RESULTS}EXPERIMENTAL RESULTS}

Fig. 2(a) shows typically measured $\delta T$ and $\delta L$ oscillations
($T_\mathrm{0}\simeq 130$ K). The phase of these two oscillations are perfectly
aligned, leading to a linear line in Fig. 2(b) whose slope is the thermal expansion
coefficient $\alpha$. $\alpha$ can be measured using lock-in technique by
separating $\delta L$ oscillation into in-phase and out-of-phase
components in reference to $\delta T$. The in-phase component of
$\alpha$ is finite and its out-of-phase component remains nearly zero at
temperatures away from $T_\mathrm{N}$, i.e. $T>136.5$ K or $T<135$ K, suggesting that the
sample is at equilibrium and have uniform phase. Besides the AC differential measurement of $\alpha$, we can also
measure the sample's DC thickness change directly from the feedback
voltage $V_F$ through $\Delta L=GV_F$, where $G=|{\lambda}/{\pi}\cdot {d\Phi}/{dV_F}|$ is the feedback
gain.

In Fig. 2(c), $\Delta L$ taken from sample S1 ($L\simeq 430 \mu$m)
exhibits a jump of about $\Delta L/L\simeq1.5\times 10^{-4}$ at $T_\mathrm{N}=135.67$ K where the in-phase
component of $\alpha$ has a huge negative peak. This first-order phase
transition is consistent with our characterization of resistivity showed in left inset in Fig. 2(c) and previous reports that the sample makes a
transition from a high-temperature PM phase to a low-temperature AFM
phase \cite{RN221}. Unlike the smooth and gradual increase on the low temperature
side of $T_\mathrm{N}$, $\alpha$ exhibit a clear kink in Fig. 2(c) at
$T_\mathrm{S}\simeq 136.44$ K, signaling the second order phase
transition. This is consistent with the work by M. G. Kim
\emph{et al.} where a structural
transition from tetragonal to orthorhombic lattice is
observed by high resolution
X-ray diffraction studies \cite{RN221}.

\begin{figure}[!htbp]
  \includegraphics[width=0.48\textwidth]{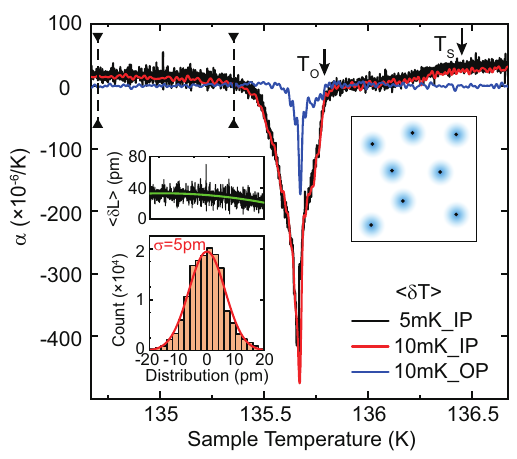}
  \caption{$\alpha$ measured using extremely small $\langle \delta T \rangle =$5mK and 10mK at 250 mHz from S2 (IP and OP are in-phase and out-of-phase part of $\alpha$, respectively). The abrupt $\alpha$ drop at $T_\mathrm{O}\simeq T_\mathrm{N}+0.1$ K suggests the formation of the AFM domains (blue dots) in the PM bath, illustrated by the right inset cartoon. The black dots represent the condensation nuclei of AFM domains, which might be defects. The left inset shows the measured $\langle \delta L \rangle$ in the range marked by the two vertical bars using 50 mHz resolution bandwidth. We deduce 5 pm resolution from the histogram of $\langle \delta L \rangle$ noise by subtracting the fitted green curve.}
\end{figure}

The small $\langle \delta T\rangle$ provides high resolution in
temperature and reveals many features near the phase
transition. Fig. 3 shows $\alpha$ measured from sample S2 ($L\simeq 210 \mu$m)
where $\langle \delta T\rangle$ is as small as 5 and 10 mK. The $\alpha$ peak of the S2 data is narrow and deep, evidencing the sample's high quality since most of its
bulk have the same $T_\mathrm{N}$. The temperature resolutions are about 10mK and 20mK and both of them are smaller than the detialed features of the $\alpha$ peak. Clear kinks in $\alpha$ appear at a
temperature $T_\mathrm{S}$ about 0.8 K above $T_\mathrm{N}$, similar to sample S1, evidencing a strong link
between these two transitions. Besides, we notice an extra abrupt change of $\alpha$ at $T_O\simeq T_\mathrm{N}+0.1$ K, which can be qualitative explained by the formation of AFM clusters around condensation nuclei such as defects while
the substantial part of the sample remains PM, see the Fig. 3 inset \cite{RN219}.

\begin{figure}[!htbp]
  \includegraphics[width=0.48\textwidth]{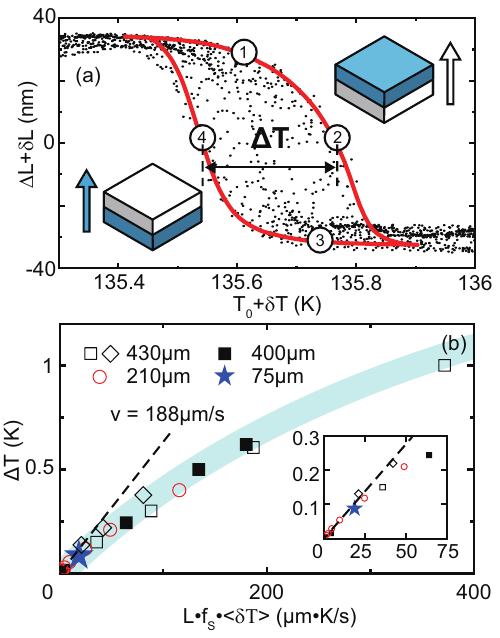}
  \caption{(a) The thickness-temperature hysteresis. Data measured using $f_S=100$ mHz and $\langle \delta T\rangle =0.25$ K. Note that the peak-to-peak amplitude of temperature oscillation is $2\langle \delta T\rangle =0.5$ K. We highlight one thermal cycle with thick red line and mark four positions in the loop. The inset cartoons illustrate the coexistence of the PM (white) and AFM (blue) phases and the propagation of their phase boundary. (b) Summarized $\Delta T$ vs. $L f_S \langle\delta T\rangle$ measured from different samples. We use $f_S=100$ (open diamond) and 250 mHz (open square) for sample S1 data, and $f_S=250$ mHz for all other data. The uncertainty of is comparable to the symbol size. The inset is a zoon-in plot near zero.}
\end{figure}

Interestingly, the phase of $\delta L$ and $\delta T$ oscillation is no longer aligned near $T_\mathrm{N}$, signaled by the large out-of-phase component of $\alpha$ in Fig 2 and Fig 3. In another word, $\delta L$ vs. $\delta T$ exhibits a hysteric ellipse, as shown Fig. 2(c) right inset. We choose the data of sample S1 at $f_S$=100 mHz and $\langle \delta T \rangle\simeq 0.25$ K as an example and sum the DC component $\Delta L$, induced by DC temperature $T_\mathrm{0}$ sweeping and measured by $V_F$, and the AC component $\delta L$, induced by AC temperature change $\delta T$ and measured by lock-in technique, as the real-time thickness of sample $\Delta L + \delta L$ and plot it as a function of the real-time
temperature $T_\mathrm{0}+\delta T$ with black dots in Fig. 4(a). We hightlight the relationship between $\Delta L+\delta L$ and $T_\mathrm{0}+\delta T$ within the one period of $\delta T$ at $T_\mathrm{0}=T_\mathrm{N}$ with red line and all the real-time thickness dots at phase transition region fall inside the area enclosed by this red line. The clear square hysteresis
loop indicates that the sample is off-equilibrium and has two coexisting phases at the transition.

A detailed and accurate description of this hysteresis involves the dynamic
process of the phase transition which is rather complicated and beyond the scope
of this article. Fortunately, we can understand the observed phenomenon using a simple toy model. In Fig. 4(a) we highlight one thermal loop with red
line and the four numbers mark four different conditions. The sample starts from a uniform AFM phase at low temperature. When the temperature increases through
$T_\mathrm{N}$ at spot 1, the PM
phase appears at the sample's bottom surface which is thermally
anchored to the sample holder. The PM domain grows and the phase boundary propagates
upward until it reaches the samples' top surface. When we cool the
sample back through $T_\mathrm{N}$ at spot 3, the AFM phase forms at the sample
bottom and grows upward. We can identify two specific positions
labeled 2 and 4 in Fig. 4(a), which correspond to the midway of the
phase transition. At these two points, the sample is divided
half-and-half into AFM and PM phases, see the inset cartoons. We
define the distance between the two points, $\Delta T\simeq 0.22$ K,
as the width of the hysteresis loop.

It is worthwhile to mention several facts about $\Delta T$. Firstly, the hysteresis width is not limited by the range of temperature oscillation, i.e. its the peak-to-peak amplitude $2\langle\delta T\rangle$. For example, $\Delta T\simeq 0.22$ K in Fig. 4(a) is about 40\% of $2\langle\delta T\rangle\simeq 0.5$ K. This ratio becomes even smaller with slower frequency or thinner sample. Secondly, in contrast with systems such as supercooled pristine water which has liquid configuration well below its crystallization temperature if cooled slowly \cite{RN233}, the finite-width hysteresis loop is only because the temperature is changing ``too fast'' for this phase transition. $\Delta T$ vanishes
when the amplitude or the frequency of $\delta T$ approaches zero.  The ratio of out-of-phase part and in-phase part of $\alpha$ can symbolize $\Delta T$ qualitatively. So the extremely narrow peak of out-of-phase part of $\alpha$ at $\langle\delta T\rangle$ =10 mK showed in Fig. 3 comparing the width peak at $\langle\delta T\rangle$ =0.14 K showed in Fig. 2(c) also indicates the transition becomes
infinitely sharp if we sweep the AC temperature change sufficiently slow, which is consistent with previous capacitive dilatometry studies where no hysteresis is seen.

We measure the hysteresis loop from different samples using different
$\delta T$ frequencies $f_S$ and amplitudes $\langle\delta
T\rangle$. We use $f_S=100$ and 250 mHz to measure S1, and $f_S=250$
mHz for the other samples. We find that $\Delta T$ is proportional to the sample
thickness $L$, the AC temperature oscillation frequency $f_S$ and
amplitude $\langle\delta T\rangle$. We summarize $\Delta T$ as a
function of $L\cdot f_S \langle\delta T\rangle$ in Fig. 4(b). It is quite
remarkable that all data points collapse onto the same curve
highlighted by the blue band. We
note that the phase boundary propagates by a distance $L/2$ to the midway of the sample
from its bottom surface at spots 2 and 4 as shown in the cartoons of Fig. 4(a), and $f_S
\langle\delta T\rangle$ is the maximum temperature changing rate. Therefore, the
slope of the blue curve corresponds to the propagation speed of phase
boundary which is $v=188$
$\mu$m/s at small $L\cdot f_S \langle\delta T\rangle$. This speed
is seven orders of magnitude
lower than that of acoustic waves \cite{RN234}.
The constraint on the phase boundary propagation speed is likely related to magnetoelastic nature of the transition at $T_\mathrm{N}$ where the electronic degrees of freedom correlates with the lattice deformation. The formation of structural domains\cite{RN247} is not directly related to the hysteresis since the domains span the whole sample along $c-$axis direction. However, it may be one of the reason limiting the propagation speed by causing strain and dissipiting energy.

\section{\label{CONCLUSION}CONCLUSION}
Our study of thermal expansion coefficient of
BaFe$_2$As$_2$ using an interferometer-based dilatometer reveals
interesting information of its magnetic transition. Our results
clearly resolve the two-step transition where the second-order
structural transition appears at $T_\mathrm{S}$ and the first-order magnetic
transition at $T_\mathrm{N}$. Thanks to the extremely high
resolution and the "true" differential nature of our technique, we discover 
the samples' thickness-temperature hysteresis loop at $T_\mathrm{N}$. We can describe this
dynamic process with a simple model and our systematical study reveals a propagation speed
of phase boundary to be $v=188$ $\mu$m/s. Further understanding for the intrinsic origin of the phase boundary propagation speed needs more relevant research. This work highlights that
AC-temperature dilatometry with extraordinarily high resolution is
a powerful probe of correlation effects in quantum materials and it's a completely new high resolution approach for phase transition research.

\begin{acknowledgments}

  The work at PKU was supported by the National Key Research and Development Program of China (Grant No. 2021YFA1401900 and 2019YFA0308403), the Innovation Program for Quantum Science and Technology (Grant No. 2021ZD0302602), and the National Natural Science Foundation of China (Grant No. 92065104 and 12074010). The work at IOP was supported by the National Key Research and Development Program of China (Grants No. 2022YFA1403400 and 2021YFA1400400), and the Strategic Priority Research Program(B) of the Chinese Academy of Sciences (Grants No. XDB33000000 and No. GJTD-2020-01). We thank Mingquan He for valuable discussion.
  
\end{acknowledgments}

\bibliographystyle{apsrev4-1}
\bibliography{references.bib}

\end{document}